\documentclass{article}

\usepackage{arxiv}

\usepackage[utf8]{inputenc} 
\usepackage[T1]{fontenc}    
\usepackage{hyperref}       
\usepackage{url}            
\usepackage{booktabs}       
\usepackage{amsfonts}       
\usepackage{nicefrac}       
\usepackage{microtype}      
\usepackage{lipsum}

\usepackage{graphicx}
\usepackage{dcolumn}
\usepackage{bm}

\usepackage{amsmath}

\usepackage{mathrsfs}
\usepackage{tikz}
\usepackage{amssymb}

\usepackage{graphicx}

\usepackage[algoruled,vlined]{algorithm2e}

\newcommand{\DDg}[4]{{#1}_{#2}^{#3}(#4)} 
\newcommand{\DD}[3]{{#1}_{#2}^{#3}}

\title{Recursive Formula for Labeled Graph Enumeration}

\author{
  Ravi Goyal\\
  Mathematica \\
  600 Alexander Park, Suite 100\\
  Princeton, NJ 08540\\
  \texttt{rgoyal@mathematica-mpr.com} \\
   \And
  Victor De Gruttola \\
  Department of Biostatistics\\
  Harvard T.H. Chan School of Public Health\\
  655 Huntington Avenue \\
  Boston, MA 02115\\
  \texttt{degrut@hsph.harvard.edu} \\
}

\begin{document}
\maketitle

\begin{abstract}
This manuscript presents a general recursive formula to estimate the size of fibers associated with algebraic maps from graphs to summary statistics of importance for social network analysis, such as number of edges (graph density), degree sequence, degree distribution, mixing by nodal covariates, and degree mixing. That is, the formula estimates the number of labeled graphs that have given values for network properties. The proposed approach can be extended to additional network properties (e.g., clustering) as well as properties of bipartite networks. For special settings in which alternative formulas exist, simulation studies demonstrate the validity of the proposed approach. We illustrate the approach for estimating the size of fibers associated with the Barab\'{a}si--Albert model for the properties of degree distribution and degree mixing. In addition, we demonstrate how the approach can be used to assess the diversity of graphs within a fiber. 
\end{abstract}


\section{Introduction}

The disciplines of algebra and combinatorics  offer critical approaches and constructs for conducting statistical network analysis. Of particular interest for this paper are algebraic maps, denoted as $\phi$, from graphs to summary statistics as well as the inverse images associated with these maps, denoted as $c_{\phi}(x) = \{g : \phi(g) = x, g \in \mathscr{G}_n\}$ where $\mathscr{G}_n$ is the set of all simple networks with $n$ vertices. These inverse images of singleton sets have been referred to as fibers in algebraic statistics literature \cite{petrovic2017survey}. We use $\phi$ to represent the mapping as well as the associated network property being calculated. Fibers have been shown to be a useful concept in social network analysis; for example, they have been used as reference sets to conduct exact tests for goodness-of-fit of exponential random graph models  \cite{gross2017goodness}. The focus of this paper is on estimating the size of a fiber, denoted as $\vert c_{\phi}(x) \vert$, which represents the number of graphs where network property $\phi$ equals $x$; this quantity has been referred to as a volume factor \cite{Shalizi13}.

Understanding how frequently a particular network feature occurs in a set of potential graphs is important for several types of social network analyses. For example, knowing the volume factor can be necessary to calculate the probability distribution on $\mathscr{G}_n$ when  network property values are known or can be estimated, but no data is known about individual edges. This scenario occurs in the analysis of sampled social network data--particularly in the setting of sexual contact networks. An example arose in the design of the Botswana Combination Prevention Program, a large cluster randomized trial of HIV prevention. For guiding the design, only estimates were available for crucial network properties, such as degree distribution (proportion of people with given numbers of sexual contacts) and proportion of sexual contacts that occur between residents of the same community  \cite{wang2014sample}. From knowledge of $\vert c_{\phi}(x) \vert$, however, we are able to estimate the probability of any given graph $g \in \mathscr{G}_n$ as $\vert c_{\phi}(\phi(g)) \vert^{-1}$ multiplied by the probability estimate for $\phi(g)$. Such knowledge permits simulation of processes operating on networks and the impact of interventions on them--which can be useful to investigate the suitability of different study designs. 

In addition, the ability  to estimate volume factors enables investigation of the diversity of graphs in a fiber. For example, a fiber based on $\phi$ being only the number of edges in a graph might have a large number of unique degree distributions (high diversity) or few (low diversity). Knowledge about the size and diversity of a fiber can aid in understanding mixing times for MCMC schemes to sample graphs from the fiber (small size or low diversity can result in faster mixing times). Also, knowledge about the diversity can aid in the assessment of the generalizability of simulation studies based on networks sampled from a fiber (higher diversity may suggest greater generalizability of study results).   

Graph enumeration is a well-established area of combinatorics for counting networks with a particular network feature. Processes for counting fall into two categories based on whether vertices in the graph are labeled or unlabeled. In the former, the vertices of the graph are labeled in a way that makes them distinguishable from one another. In the latter, all permutations of the vertices are considered to form the same graph. In social network analysis--our area of interest--vertices are most often distinguishable from each other; hence, we focus on labeled graph enumeration. 

Graph enumeration dates back at least to 1889 when Cayley provided an equation to count the number of labeled trees--connected graphs that contains no cycles \cite{cayley1889theorem}. Equations to calculate the number of labeled graphs with various characteristics have since been reported; these include  rooted graphs, connected graphs, and directed graphs. Considerable research has been devoted to estimating the number of labeled graphs with a given degree sequence--a property important in social network analysis \cite{read1959enumeration, bender1978asymptotic, bollobas1980probabilistic, liebenau2017asymptotic}. However, there has been little research focusing on other important properties in social network analysis, such as degree mixing and clustering. Harary and Palmer \cite{harary2014graphical} provide an excellent introduction on graph enumeration. 

This paper presents a general recursive formula to estimate the number of labeled graphs for given values of graph properties of relevance to social network analysis: number of edges (graph density), degree sequence, degree distribution, mixing by nodal covariates, and degree mixing. The next section presents a general recursive formula to estimate the number of labeled graphs as well as details to evaluate the formula for specific network properties. For settings in which alternative formulas exist, section 3 presents simulation studies demonstrating the degree of similarity among these methods. In section 4, we apply the proposed approach to estimate the number of labeled graphs associated with different values of degree distribution and degree mixing that arise from the Barab\'{a}si--Albert model in order to investigate the diversity within the fibers associated with this model \cite{BA99}. The paper concludes with a discussion and further research.

\section{Recursive formula for graph enumeration}

We represent a network, $g = (V, E)$, as an adjacency matrix with dimensions equal to the size of set $V$. Therefore, $g$ has dimensions $\vert V \vert \times \vert V \vert$, where $\vert V \vert$ denotes the size of set $V$. Let $n$ represent the number of vertices in $g$, i.e., $n = \vert V \vert$. Let $\{v_1,\cdots,v_{n}\}$ denote the vertices in set $V$. Let $g[i,j] = 1$ indicate that there is an edge between $i$ and $j$, where $i, j \in \{v_1,\cdots, v_{n}\}$, while $g[i,j] = 0$ indicates that there is no edge and denote the neighbors of $i$ as $\eta(i)$, i.e., $\eta(i) = \{j : g[i,j] = 1\}$.

Equation~\ref{eq:recursive_formula} provides a recursive formula to estimate the number of graphs, $\vert c_{\phi}(x_k) \vert$, with specific value(s), $x_k$, for particular network properties, $\phi$:

\begin{equation}\label{eq:recursive_formula}
\vert c_{\phi}(x_k) \vert = r_{\phi}(x_k, x_{k-1}) * \vert c_{\phi}(x_{k-1}) \vert, 
\end{equation}

\noindent where $r_{\phi}(x_k, x_{k-1})$ is the ratio between the sizes of fibers $c_{\phi}(x_{k})$ and $c_{\phi}(x_{k-1})$, i.e.,

\begin{equation}\label{eq:recursive_formula_ratio}
r_{\phi}(x_k, x_{k-1}) = \frac{\vert c_{\phi}(x_{k}) \vert }{\vert c_{\phi}(x_{k-1}) \vert}.
\end{equation}

Goyal et al. \cite{goyal2014sampling} provides equations to calculate $r_{\phi}(x_i, x_{i-1})$ for a range of network properties including number of edges, mixing by nodal covariates, degree distribution, degree mixing, and clustering when $x_i$ and $x_{i+1}$ are specified such that there exists graphs $g_i$ and $g_{i-1}$ where:

\begin{enumerate}
\item[{s1.}] $g_i$ and $g_{i-1}$ differ by the presence or absence of a single edge,
\item[{s2.}] $\phi(g_i) = x_{i}$, and
\item[{s3.}] $\phi(g_{i-1}) = x_{i-1}$;
\end{enumerate}

Although there is no constraint on $x_0$, it is often useful to set $x_0$ equal to the specific value of the network properties associated with the empty graph; hence, typically,  $\vert c_{\phi}(x_0) \vert$ = 1. 

In the sections below, we provide details for calculating $\vert c_{\phi}(x) \vert$ for a given number of edges, degree distribution, degree sequence, mixing by nodal covariates, and degree mixing.  We make use of  $s1-s3$ for specifying $\{x_0,\cdots,x_k\}$ and set  $x_0$ equal to the specific value of the network properties associated with the empty graph.

\subsection{Number of edges}

Although there is a closed-form expression for calculating the number of labeled graphs with the number of edges equal to $x$, we use Equation~\ref{eq:recursive_formula} for this calculation to illustrate  its use in  graph enumeration. 

Let $\phi_1$ denote the network property for the number of edges. Specify $x_0, \cdots, x_k=x$ as $x_i = i$. As only the empty graph has $0$ edges, $\vert c_{\phi_1}(x_0) \vert = 1$. To prove that this specification satisfies $s1-s3$, let $E=\{e_1,\cdots,e_k\}$ be a set of distinct edges. Let $g_i$ denote the network formed with the first $i$ edges from $E$, i.e., $g_i$ contains edges $\{e_1,\cdots,e_i\}$. Based on the definition of $g_i$, $\phi_1(g_i) = x_i$, $\phi(g_{i-1}) = x_{i-1}$, and $g_i$ and $g_{i-1}$ differ by a single edge. Since $x_0, \cdots, x_k=x$ satisfies $s1-s3$, we can use results from Goyal et al. \cite{goyal2014sampling} to calculate $r_{\phi_1}(x_i, x_{i-1})$ as shown below:

\begin{equation}\label{eq:recursive_formula_ratio_edges}
r_{\phi_1}(x_i, x_{i-1}) = \frac{ \binom{n}{2} - x_{i-1} }{x_{i}}.
\end{equation}

Using Equation~\ref{eq:recursive_formula_ratio_edges} along with the specification of $x_0, \cdots, x_k=x$ as $x_i = i$ and $\vert c_{\phi_1}(x_0) \vert = 1$, it is possible to calculate $\vert c_{\phi_1}(x) \vert$. Section 4.1 provides a comparison between the recursive formula and a previously established formula. 

\subsection{Degree distribution}

The degree of vertex $i$, denoted as $\DDg{d}{i}{}{g}$, is the number of edges the vertex has with other vertices in $V$; therefore $\DDg{d}{i}{}{g} = \sum_j g[i,j]$. Let $\DDg{d}{}{}{g}= (\DDg{d}{1}{}{g},\cdots, \DDg{d}{n}{}{g})$ represent the vector of degrees for nodes in set $V$, commonly referred to as a degree sequence. The degree distribution, denoted as $\DDg{D}{}{}{g}$, is a vector representing the number of these degrees over all vertices in set $V$; the $k^{th}$ entry represents the number of vertices having degree $k$, i.e., $\DDg{D}{k}{}{g} = \sum_{i=1}^{n_t} I_{\{\DDg{d}{i}{}{g} = k\}}$. Let $\phi_2$ denote the network property for the degree distribution. 

To calculate $\vert c_{\phi_2}(x) \vert$, the number of labeled graphs with degree distribution $x$, we specify $x_0,\cdots,x_k=x$ by leveraging the Havel--Hakimi algorithm. Let $d = (d_1,..,d_n)$ be any degree sequence that is consistent with $x$. The Havel--Hakimi algorithm permits identification of  a set of edges, denoted as $E$, that can be used to construct a graph with degree sequence $d$ \cite{SLH62, VH55}.  Algorithm 1, provides a procedure to identify $E$.

\begin{algorithm}
  \caption{Degree distribution}
  Part 1: Generate degree sequence\;
  $S \gets \emptyset$\;
  \For{$j \gets 0$ \KwTo $n$}
  {
    $S \gets \text{Union(S,Rep(j,x[j]))}$\;
  }
  Part 2: Generate edges\;
  $E \gets \emptyset$\;
  \For{$i \gets 1$ \KwTo $n$}
  {
    $v \gets \text{Index of maximum elt. in S}$\;
    $l \gets S[v]$\;
    $S[v] \gets 0$\;
    $v_1,\cdots,v_l \gets \text{Indices of the l maximum elts. in S}$\;
    \For{$j \gets 1$ \KwTo $l$}
    {
      $S[v_j] \gets S[v_j] - 1$\;
      $E \gets Union(E, (v, v_j))$\;
    }
  }
  \KwRet{$E$}\;
\end{algorithm}

Let $g_i$ denote the network formed with the first $i$ edges from $E$, i.e., contains edges $\{e_1,\cdots,e_i\}$. Let $x_i$ denote the degree distribution associated with network $g_i$, i.e., $x_i = \DDg{D}{}{}{g_i}$. Based on this definition of $g_i$, $\vert c_{\phi_2}(x_0) \vert = 1$ as only the empty graph has degree distribution $x_0$. In addition, $x_0, \cdots, x_k=x$ satisfies $s1-s3$. Let $(l,j)$ be the single edge that differs between $g_i$ and $g_{i-1}$. Based on results from Goyal et al. \cite{goyal2014sampling}:

\begin{equation}\label{eq:recursive_formula_ratio_deg_dist_1}
r_{\phi}(x_i, x_{i-1}) = \frac{ \beta_1(g_{i-1}) - \alpha_1(g_{i-1})}{ \alpha_1(g_{i})},
\end{equation}

\noindent where,

\begin{equation}\label{eq:recursive_formula_ratio_deg_dist_2}
\alpha_1(g) = E(\DD{DMM}{\DDg{d}{l}{}{g},\DDg{d}{j}{}{g}}{} \vert \DDg{D}{}{}{g});
\end{equation}

\begin{equation}\label{eq:recursive_formula_ratio_deg_dist_3}
\beta_1(g) = \left\{ \begin{gathered}
\DDg{D}{\DDg{d}{l}{}{g}}{}{g}  \DDg{D}{\DDg{d}{j}{}{g}}{}{g} 
\mbox{ if }\DDg{d}{l}{}{g} \neq \DDg{d}{j}{}{g} \hfill \\
\binom{\DDg{D}{\DDg{d}{l}{}{g}}{}{g}}{2} \mbox{ else, } \hfill \\
\end{gathered} \right.\end{equation}

\noindent and based on Newman \cite{MN02},

\begin{equation}\label{eq:recursive_formula_ratio_deg_dist_4}
E(\DD{DMM}{x,y}{} \vert \DD{D}{}{}) \approx \frac{x\DD{D}{x}{} \times y\DD{D}{y}{}}{.5(\sum_z z\DD{D}{z}{})} \times \left( \frac{1}{2} \right)^{I\{x = y\}}. 
\end{equation}

Using Equation~\ref{eq:recursive_formula_ratio_deg_dist_1} along with the specification of $x_0, \cdots, x_k=x$ as defined above and $\vert c_{\phi_2}(x_0) \vert = 1$, it is possible to calculate $\vert c_{\phi_2}(x) \vert$. 

\subsection{Degree sequence}

Let $\phi_{2a}$ denote the network property for the degree sequence. The number of graphs with degree sequence $d$, $\vert c_{\phi_{2a}}(d) \vert$, can be computed by dividing the number of labeled graphs with the degree distribution consistent with $d$, denoted as $\DDg{D}{}{}{d}$, by the number of permutations of assigning vertices to degrees. Specifically, 

\begin{equation}
\vert c_{\phi_{2a}}(d) \vert = \frac{\vert C_{\phi_2}(\DDg{D}{}{}{d}) \vert}{\prod_{j=0}^n {n-\sum_{k=0}^{j-1} \DDg{D}{k}{}{d} \choose \DDg{D}{j}{}{d}}}.
\end{equation}

Section 4.2 provides a comparison between the presented recursive formula and a formula by Liebenau et al. \cite{liebenau2017asymptotic}. 

\subsection{Mixing by nodal covariates}

Let $\DDg{m}{i}{}{g}$ represent the vector of discrete characteristics for individual $i$ in network $g$. Let $\DDg{m}{}{}{g} = (\DDg{m}{1}{}{g},\cdots, \DDg{m}{n}{}{g})$ be a vector containing the characteristics of all vertices. The characteristic distribution, denoted as $\DDg{M}{}{}{g}$, is a vector representing the number of individuals with these characteristics over all vertices; the $k^{th}$ entry represents the number of vertices having characteristic pattern $k$, i.e., $\DDg{M}{k}{}{g} = \sum_{i = 1}^{n} I_{\{\DDg{m}{i}{}{g} = k\}}$. Let $\DDg{MM}{}{}{g}$ be a matrix representing the mixing pattern of network $g$. $\DDg{MM}{}{}{g}$ is a $q \times q$ symmetric matrix, where $q$ is the number of distinct characteristic patterns. The entry $\DDg{MM}{k,l}{}{g}$ is the total number of edges between a vertex with characteristic $k$ and vertex with characteristic $l$. Let $\phi_{3}$ denote the network property for mixing by nodal covariates. 

To calculate the number of labeled graphs with mixing matrix $x$, specify $x_0, \cdots, x_k=x$ as the following for $l leq m$ ($x_i$ is symmetric): 

\begin{equation}
\DD{x}{i_{l,m}}{} = \left\{ \begin{gathered}
0 \mbox{ if } i \leq \sum_{a=1}^{l-1} \sum_{b=a}^{q} \DD{x}{a,b}{} + \sum_{b=l}^{m-1} \DD{x}{l,b}{} \hfill \\
\DD{x}{l,m}{} \mbox{ else if } i \geq \sum_{a=1}^{l-1} \sum_{b=a}^{q} \DD{x}{a,b}{} + \sum_{b=l}^{m} \DD{x}{l,b}{}  \hfill \\
i - \sum_{a=1}^{l-1} \sum_{b=a}^{q} \DD{x}{a,b}{} + \sum_{b=l}^{m-1} \DD{x}{l,b}{} \mbox{ else, } \hfill \\
\end{gathered} \right.
\end{equation}

\noindent where $q$ is the number of distinct characteristic patterns. Therefore, $\vert c_{\phi_3}(x_0) \vert = 1$ as only the empty graph has $0$ for all entries of the mixing matrix. Also, this specification satisfies $s1-s3$. To show this, let $E=\{e_1,\cdots,e_k\}$ be a set of distinct edges where the first $\DD{x}{1,1}{}$ are between vertices with covariate pattern $1$, the next $\DD{x}{1,2}{}$ are between vertices with covariate patterns $1$ and $2$, and so on. Let $g_i$ denote the network formed with the first $i$ edges from $E$, i.e., $g_i$ contains edges $\{e_1,\cdots,e_i\}$. Based on the definition of $g_i$, $\phi_3(g_i) = x_i$, $\phi_3(g_{i-1}) = x_{i-1}$, and $g_i$ and $g_{i-1}$ differ by a single edge; let $(l,j)$ be that edge. Therefore, $r_{\phi_3}(x_i, x_{i-1})$ is the following:\\

\noindent If $m_l \neq m_j$

\begin{equation}  \label{eq:recursive_formula_ratio_mixing_1}
r_{\phi_3}(x_i, x_{i-1}) = 
    \frac{\DD{M}{m_l}{} \times \DD{M}{m_j}{}  - \DD{x}{i-1_{l,j}}{}}
    {\DD{x}{i_{l,j}}{}},
\end{equation}

\noindent else,

\begin{equation} \label{eq:recursive_formula_ratio_mixing_2}
r_{\phi_3}(x_i, x_{i-1}) =   
    \frac{{\DD{M}{m_l}{} \choose 2} - \DD{x}{i-1_{l,j}}{}}
    {\DD{x}{i_{l,j}}{}},
\end{equation}

Using Equations~\ref{eq:recursive_formula_ratio_mixing_1} and~\ref{eq:recursive_formula_ratio_mixing_2} along with the specification of $x_0, \cdots, x_k=x$ as defined above and $\vert c_{\phi_3}(x_0) \vert = 1$, it is possible to calculate $\vert c_{\phi_3}(x) \vert$. 

\subsection{Degree mixing}

Let $\DDg{DMM}{}{}{g}$ be a matrix representing the degree mixing pattern of network $g$. The entry $\DDg{DMM}{k,l}{}{g}$ is the total number of edges between vertices of degree $k$ and $l$. Let $\phi_{4}$ denote the network property for degree mixing matrix. 

To calculate the number of labeled graphs with degree mixing $x$, we follow a similar approach as that for degree distribution. Specifically, we use a constructive proof for assessing whether a degree mixing matrix is graphical in order to specify a set of edges, $E$, that can be used to construct a graph with degree mixing $x$ \cite{goyal2014sampling}; Algorithm 2 provides a procedure to construct $E$. 

\begin{algorithm}
  \caption{Degree Mixing}
  Part 1: Generate degree distribution\;
  \For{$j \gets 0$ \KwTo $n-1$}
  {
    $\DD{D}{j}{} \gets (\sum_{i=0} \DD{DMM}{i,j}{} + \DD{DMM}{j,j}{}) / j$\;
  }
  Part 2: Generate degree sequences\;
  $S_{c} \gets Rep(0, n)$\;
  $S_{f} \gets \emptyset$\;
  \For{$j \gets 0$ \KwTo $n$}
  {
    $S_f \gets \text{Union(}S_{f}\text{,Rep(j,}\DD{D}{j}{}\text{))}$\;
  }
  Part 3: Edges between vertices with same final degree\;
  $E \gets \emptyset$\;
  \For{$j \gets 0$ \KwTo $n$}
  {
    $P \gets {0,\cdots,2*\DD{DMM}{j,j}{}}-1 \text{ modulo } \DD{D}{j}{}$\;
    $T \gets \text{Table(P)}$\;
    $S_j \gets \emptyset$\;
    \For{$i \gets 0$ \KwTo $n$}
    {
        \uIf{i = j}
        {
            $S_j \gets \text{Union(} S_j \text{,T)}$\;
        }
        \Else
        {
            $S_j \gets \text{Union(} S_j \text{,Rep(0,} \DD{D}{i}{} \text{))}$\;
        }
    }
    $S_{c} \gets S_{c} + S_j$\;
    $E_j \gets\text{Use degree distribution algorithm with } S_j$\;
    $E \gets \text{Union(E,}E_j\text{)}$\;
  }
  Part 4: Edges between vertices with different final degrees\;
  \For{$j \gets 1$ \KwTo $n-1$}
  {
    \For{$i \gets j+1$ \KwTo $n$}
    {
        $P \gets {0,\cdots,\DD{DMM}{j}{i}-1} \text{ modulo } \DD{D}{j}{}$\;
        $T \gets \text{Table(P)}$\;
        \For{$k \gets 1$ \KwTo $\DD{D}{j}{}$}
        {
            $v \gets \text{Index of min. elt. in } S_{c} \text{ s.t. } S_{f} = j$\;
            $w_1,\cdots,w_{T_k} \gets \text{Indices of }T_k \text{ min. elts. in } S_{c} \text{ s.t. } S_{f} = i$\;
            \For{$m \gets 1$ \KwTo $T_k$}
            {
                $S_{c}[v] \gets S_{c}[v] + 1$\;
                $S_{c}[w_m] \gets S_{c}[w_m] + 1$\;
                $E \gets Union(E, (v, w_m))$\;
            }
        }
    }
  }
  \KwRet{E}\;
\end{algorithm}

Let $g_i$ denote the network formed with the first $i$ edges from $E$, i.e., contains edges $\{e_1,\cdots,e_i\}$. Let $x_i$ denote the degree mixing associated with network $g_i$, i.e., $x_i = \DDg{DMM}{}{}{g_i}$. As with number of edges and degree distribution, $x_0, \cdots, x_k=x$ satisfies conditions $s1-s3$ and $x_0$ takes on the value associated with the empty graph. Therefore, $\vert c_{\phi_4}(x_0) \vert = 1$. Let $(l,j)$ be the single edge that differs between $g_i$ and $g_{i-1}$. Based on results from Goyal et al. \cite{goyal2014sampling}:

\begin{equation}\label{eq:recursive_formula_ratio_deg_mix_1}
r_{\phi_4}(x_i, x_{i-1}) = \frac{ [\gamma_1(g_{i-1}) - \alpha_2(g_{i-1})] \times  
\DDg{\beta}{(l,j)}{0}{g_{i-1}} }{ \DDg{DMM}{\DDg{d}{l}{}{g_i},\DDg{d}{j}{}{g_i}}{}{g_i} \times 
\DDg{\beta}{(l,j)}{1}{g_i}},
\end{equation}

\noindent where,

\begin{equation}\label{eq:recursive_formula_ratio_deg_mix_2}
\alpha_2(g) =  \DDg{DMM}{\DDg{d}{l}{}{g},\DDg{d}{j}{}{g}}{}{g};
\end{equation}

\begin{equation}\label{eq:recursive_formula_ratio_deg_mix_3}
\gamma_1(g) = \left\{ \begin{gathered}
\DDg{D}{\DDg{d}{i}{}{g}}{}{g} \DDg{D}{\DDg{d}{j}{}{g}}{}{g} \mbox{ if }  \DDg{d}{i}{}{g} \neq \DDg{d}{j}{}{g} \hfill \\
{ \DDg{D}{\DDg{d}{i}{}{g}}{}{g} \choose 2 } \mbox{ else;} \\
\end{gathered} \right.\end{equation}

\noindent and based on concepts from Newman \cite{MN02}, if $\DDg{d}{i}{}{z} \neq \DDg{d}{j}{}{z}$,

\begin{equation}
\DDg{\beta}{(l,j)}{s}{g} \approx 
\frac{\Pi_z {  \DDg{DMM}{\DDg{d}{l}{}{g},z}{'}{g}  - I_{\{\DDg{d}{j}{}{g} = z\}} \cdot s \choose n_l^z -  I_{\{\DDg{d}{j}{}{g} = z\}} \cdot s}}
{ {\sum_z  \DDg{DMM}{\DDg{d}{l}{}{g},z}{'}{g}  - I_{\{\DDg{d}{j}{}{g} = z\}} \cdot s \choose  \DDg{d}{l}{}{g} - s}} \times \frac{\Pi_z {  \DDg{DMM}{z,\DDg{d}{j}{}{g}}{'}{g}  - I_{\{\DDg{d}{l}{}{g} = z\}} \cdot s \choose n_j^z -  I_{\{\DDg{d}{l}{}{g} = z\}} \cdot s}}
{ {\sum_z  \DDg{DMM}{z,\DDg{d}{j}{}{g}}{'}{g}  - I_{\{\DDg{d}{l}{}{g} = z\}} \cdot s \choose  \DDg{d}{j}{}{g} - s}} 
\end{equation}

\noindent else, 

\begin{equation}
\DDg{\beta}{(l,j)}{s}{g} \approx 
\frac{\Pi_z {  \DDg{DMM}{\DDg{d}{l}{}{g},z}{'}{g}  - I_{\{\DDg{d}{j}{}{g} = z\}} \cdot s \choose n_l^z + n_j^z -  2I_{\{\DDg{d}{j}{}{g} = z\}} \cdot s}}
{ {\sum_z  \DDg{DMM}{\DDg{d}{l}{}{g},z}{'}{g}  - I_{\{\DDg{d}{j}{}{g} = z\}} \cdot s \choose \DDg{d}{l}{}{g} + \DDg{d}{j}{}{g} - 2s}}.
\end{equation}

\noindent where, $\DDg{DMM}{a,b}{'}{g} = \DDg{DMM}{a,b}{}{g}$ if $a \neq b$ and $\DDg{DMM}{a,b}{'}{g} = 2*\DDg{DMM}{a,b}{}{g}$ if $a = b$ and $n_l^z$ and $n_j^z$ denote the number of vertices that are neighbors of $i$ and $j$ and equal to $z$.

Using Equation~\ref{eq:recursive_formula_ratio_deg_mix_1} along with the specification of $x_0, \cdots, x_k=x$ as defined above and $\vert c_{\phi_4}(x_0) \vert = 1$, it is possible to calculate $\vert c_{\phi_4}(x) \vert$. 

\subsection{Additional network properties and bipartite graphs}

The recursive formula and associated framework we propose can be used to calculate the number of labeled graphs for many additional network properties. In particular, Goyal et al. \cite{goyal2014sampling} provide  equations for $r_{\phi}(x_k, x_{k-1})$ for clustering (controlling for degree mixing) and mixing by nodal covariates (controlling for degree distribution). In addition, Goyal et al. \cite{goyal2018inference} enables extending the calculation of $r_{\phi}(x_k, x_{k-1})$ to the setting of  bipartite networks. 

\section{Comparison with previous research}

\subsection{Number of edges}

The number of graphs of size $n$ with $x$ edges  equals the following:

\begin{equation} \label{eq:std_edge_formula}
\vert c_{\phi_1}(x) \vert = {{n \choose 2} \choose x}.
\end{equation}

\noindent This holds because  there are ${n \choose 2}$ possible edges and $x$ of those edges are selected.  

To  illustrate the use and validity of the recursive formula, we compare the estimates of $\vert c_{\phi_1}(x) \vert$ based on the proposed recursive formula to the formula in Equation~\ref{eq:std_edge_formula}; we consider values of  $x$ ranging from $\{1,\cdots,10\}$ for graphs of size $n = 1000$.        

For each value of $x \in \{1,\cdots,10\}$, Table~\ref{table:comparison_edges} provides the log values for $r(x, x-1)$ and $\vert c_{\phi}(x) \vert$ based on Equation~\ref{eq:recursive_formula} and Equation~\ref{eq:std_edge_formula}. The estimates obtained from  the proposed recursive formula and the known formula are identical--an expected finding given that a closed-form equation for $r_{\phi_1}(x,x-1)$  exists.

\begin{table}
\caption{Comparison of methods to calculate number of edges}
\centering
\begin{tabular}[t]{ccc|c}
\toprule
$x$ & $log(r(x, x-1))$ & $log(\vert c_{\phi_1}(x) \vert)$ & $log(\vert c_{\phi_1}(x) \vert)$ \\
 &  & [equation~\ref{eq:recursive_formula}] & [equation~\ref{eq:std_edge_formula}] \\
\hline
 $0$ & $-$ & $0$ & $0$\\
$1$ & $13.12$ & $13.12$ & $13.12$\\
$2$ & $12.42$ & $25.55$ & $25.55$\\
$3$ & $12.02$ & $37.57$ & $37.57$\\
$4$ & $11.74$ & $49.31$ & $49.31$\\
$5$ & $11.51$ & $60.82$ & $60.82$\\
$6$ & $11.33$ & $72.15$ & $72.15$\\
$7$ & $11.18$ & $83.32$ & $83.32$\\
$8$ & $11.04$ & $94.37$ & $94.37$\\
$9$ & $10.92$ & $105.29$ & $105.29$\\
$10$ & $10.82$ & $116.11$ & $116.11$\\
\bottomrule
\end{tabular}
\label{table:comparison_edges}
\end{table}

\subsection{Degree sequence}

Liebenau et al. \cite{liebenau2017asymptotic} proved a general asymptotic formula--conjectured in 1990--for the number of graphs with given degree sequence. They also provide  a formula that converges to the  number of d-regular graphs for any values of d as $n\to\infty$; a d-regular graph is one in which each vertex has exactly degree $d$. We compare estimates of $\vert c_{\phi}(x) \vert$, where $x$ is the degree sequence $\DD{d}{}{l}=\{\DD{d}{1}{l},\cdots,\DD{d}{n}{l}\}$ and $\DD{d}{i}{l} = l$ for all $i \in \{1,\cdots,n\}$, obtained from this asymptotic formula to those from the proposed recursive formula.   Our estimates in this section are based on $x_0$ being the degree sequence $\DD{d}{}{0}$.

\begin{figure*}
\includegraphics[scale=1.05]{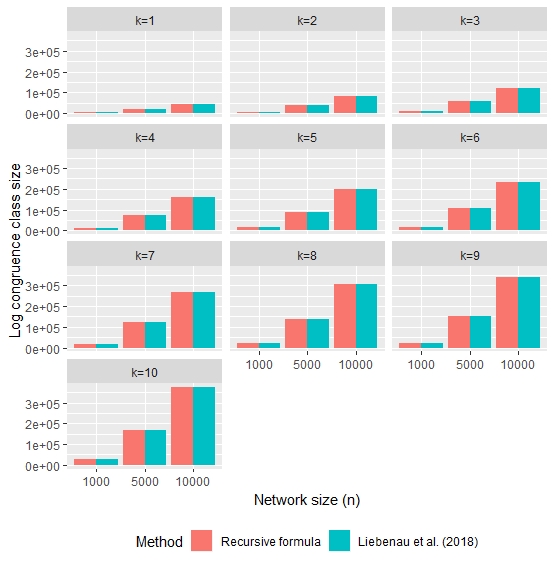}
\caption{\label{fig_deg_distr_1}\textbf{Comparison:} Log estimates for $k \in \{1,\cdots,10\}$ for the two approaches. The red bars depict estimates based on the recursive formula introduced in this manuscript; the blue bars are estimates based on Liebenau et al. \cite{liebenau2017asymptotic}. Each plot shows log estimates for networks of size 1000, 5000, and 10000.}
\end{figure*}

Figure~\ref{fig_deg_distr_1} shows log estimates for $k \in \{1,\cdots,10\}$ for the two approaches. The red bars depict estimates based on the recursive formula introduced in this manuscript; the blue bars are estimates based on Liebenau et al.  \cite{liebenau2017asymptotic}. Each plot in Figure~\ref{fig_deg_distr_1} shows log estimates for networks of size 1000, 5000, and 10000.  The log estimates differ by less than $0.01\%$.

\section{Fiber sizes for the Barab\'{a}si--Albert model}

In this section, we estimate the fiber sizes associated with degree distributions and degree mixing matrices when graphs are generated using the Barab\'{a}si--Albert (BA) model. In addition, we estimate fiber sizes of graphs that are generated under different models, but constrained to share specific values for network properties with the graphs from the BA. We compare these fiber sizes to investigate the sizes of the fibers associated with graphs generated from the BA model relative to the other models--specifically the Erd\H{o}s-R\'enyi (ER) model and configuration (CONF) model. Finally, we estimate the diversity in degree mixing matrices consistent with degree distributions formed by the BA model. First, we provide details of the BA model.

The BA model can be initiated with a small seed graph that  grows  by the addition of  new vertices one at a time. Each new vertex forms a new edge with an existing vertex based on preferential attachment rules. Vertices and edges, once introduced, are never deleted. The BA model fixes the number of (undirected) edges connected to each new vertex. (Note that the model can be  modified in various ways; our focus here is conceptual, so we consider only the original versions of these models.) The BA model provides one mechanism to generate graphs with a fat-tailed degree distribution, specifically a power-law degree distribution, where the probability, $P(k)$, that a vertex in the graph has degree $k$ decays as a power-law $P(k) \sim k^{-\gamma}$. By contrast, the degree distributions for the ER model follow a binomial distribution.  
\subsection{Comparison of the BA and ER models}

To compare the sizes of the fibers based on degree distribution associated with the BA and ER models, we generate 100 graphs using the BA model ($n=5000$), denoted as $\{g^{BA}_1,\cdots,g^{BA}_{100}\}$, and 100 graphs using the ER model, denoted as $\{g^{ER}_1,\cdots,g^{ER}_{100}\}$, such that $\phi_1(g^{ER}_i) = \phi_1(g^{BA}_i)$. 

\begin{figure*} 
\centerline{\includegraphics[scale=.6]{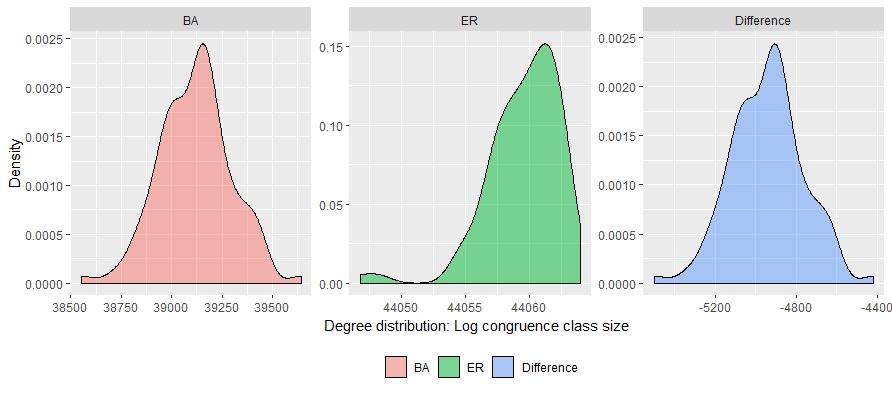}}
\caption{\textbf{Fiber sizes for degree distribution:} Density plots for the log estimates for $\vert c_{\phi_2}(\phi_j(g^{BA}_i)) \vert$ (first panel), $\vert c_{\phi_2}(\phi_j(g^{ER}_i)) \vert$ (second panel), and their differences (third panel).}
\label{fig_deg_distr_BA_ER}
\end{figure*}

Figure~\ref{fig_deg_distr_BA_ER} shows density plots for the log estimates for $\vert c_{\phi_2}(\phi_j(g^{BA}_i)) \vert$  in the first panel; $\vert c_{\phi_2}(\phi_j(g^{ER}_i)) \vert$, second panel; and their differences, third panel. Although $g^{BA}_i$ and $g^{ER}_i$ have the same number of edges, the number of graphs with a power law distribution for their degrees derived from a BA model is much smaller that the number of graphs wherein degrees follow a binomial distribution derived from an ER model.

\subsection{Comparison of the BA and CONF models}

The BA model also produces non-random structure in other network properties besides degree distribution; these include correlations between the degrees of connected vertices \cite{qu2015effects}. To compare the sizes of the fibers based on degree mixing between graphs generated using BA model and graphs sampled uniformly with the same degree distribution, we generate 100 graphs using the configuration model, denoted as $\{g^{CONF}_1,\cdots,g^{CONF}_{100}\}$; such that $\phi_2(g^{CONF}_i) = \phi_2(g^{BA}_i)$ \cite{MN10}. 

\begin{figure*}
\centerline{\includegraphics[scale=.6]{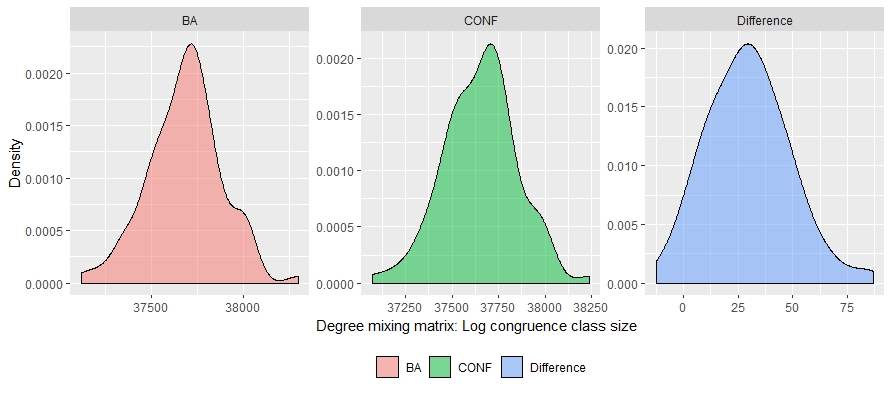}}
\caption{\textbf{Fiber sizes for degree mixing matrix:} Density plots for the log estimates for $\vert c_{\phi_4}(\phi_j(g^{BA}_i)) \vert$ (first panel), $\vert c_{\phi_4}(\phi_j(g^{CONF}_i)) \vert$ (second panel), and their difference (third panel).}
\label{fig_deg_mix_BA_CONF}
\end{figure*}

Figure~\ref{fig_deg_mix_BA_CONF} shows density plots for the log estimates for $\vert c_{\phi_4}(\phi_j(g^{BA}_i)) \vert$ (first panel), $\vert c_{\phi_4}(\phi_j(g^{CONF}_i)) \vert$ (second panel), and their difference (third panel). The number of graphs with a degree mixing matrix derived from a BA model are similar on the log scale to the number of graphs with a degree mixing matrix derived from the configuration model.

\subsection{Diversity of degree mixing matrices}

From the previous section, the average number of labeled graphs associated with a degree distribution generated from the BA model was estimated as $1.26e16988$ (exponential of the mean of the first panel in Figure~\ref{fig_deg_distr_BA_ER}). This section explores the diversity within this large collection of graphs. Specifically, we estimate the number of distinct degree mixing matrices associated with a degree distribution generated from the BA model.
     
\begin{figure}
\centerline{\includegraphics[scale=.5]{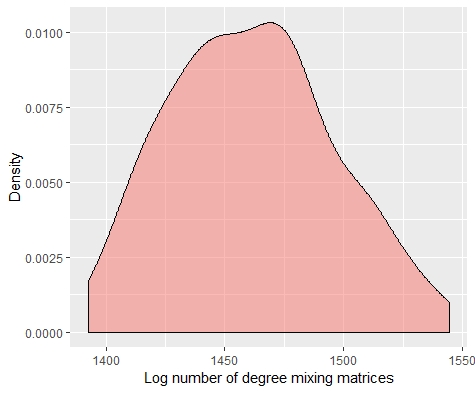}}
\caption{\textbf{Log number of degree mixing matrices:} Density plot for the log of the estimates of number of distinct degree mixing matrices associated with a degree distribution generated from the BA model.}
\label{fig_num_deg_mix}
\end{figure}

Figure~\ref{fig_num_deg_mix} shows a density plot for the log estimates for the number of distinct degree mixing matrices associated with a degree distribution generated from the BA model. The exponential of the mean gives an estimate of $4.16e634$ distinct degree mixing matrices.

\section{Discussion}

This paper presents a general recursive formula to estimate the number of labeled graphs with specific values for graph properties of interest. We consider those with particular relevance for social network analysis: number of edges (graph density), degree sequence, degree distribution, mixing by nodal covariates, and degree mixing.   The proposed method can easily be extended to additional network properties, including clustering, as well as to bipartite graphs; the formulas for Equation~\ref{eq:recursive_formula_ratio} are currently available. The proposed recursive formula differs from other available approaches for graph enumeration both in its overall approach and in the breadth of network properties that can be considered; it may be profitable to investigate the theoretical connections between the proposed method and other approaches.

\section{Acknowledgements}

This research is supported by a grant from the National Institute of Health (R37 AI-51164)

\bibliographystyle{unsrt}  
\bibliography{paper2_bib}

\begin{thebibliography}{10}

\bibitem{petrovic2017survey}
Sonja Petrovic.
\newblock A survey of discrete methods in (algebraic) statistics for networks.
\newblock {\em Algebraic and Geometric Methods in Discrete Mathematics},
  685:260--281, 2017.

\bibitem{gross2017goodness}
Elizabeth Gross, Sonja Petrovi{\'c}, and Despina Stasi.
\newblock Goodness of fit for log-linear network models: dynamic markov bases
  using hypergraphs.
\newblock {\em Annals of the Institute of Statistical Mathematics},
  69(3):673--704, 2017.

\bibitem{Shalizi13}
Cosma~Rohilla Shalizi\: and Alessandro Rinaldo.
\newblock Consistency under sampling of exponential random graph models.
\newblock {\em Annals of Statistics}, 41(2):508--535, 2013.

\bibitem{wang2014sample}
Rui Wang, Ravi Goyal, Quanhong Lei, M~Essex, and Victor De~Gruttola.
\newblock Sample size considerations in the design of cluster randomized trials
  of combination hiv prevention.
\newblock {\em Clinical Trials}, page 1740774514523351, 2014.

\bibitem{cayley1889theorem}
Arthur Cayley.
\newblock A theorem on trees.
\newblock {\em Quart. J. Math.}, 23:376--378, 1889.

\bibitem{read1959enumeration}
RC~Read.
\newblock The enumeration of locally restricted graphs (i).
\newblock {\em Journal of the London Mathematical Society}, 1(4):417--436,
  1959.

\bibitem{bender1978asymptotic}
Edward~A Bender\: and E~Rodney Canfield.
\newblock The asymptotic number of labeled graphs with given degree sequences.
\newblock {\em Journal of Combinatorial Theory, Series A}, 24(3):296--307,
  1978.

\bibitem{bollobas1980probabilistic}
B{\'e}la Bollob{\'a}s.
\newblock A probabilistic proof of an asymptotic formula for the number of
  labelled regular graphs.
\newblock {\em European Journal of Combinatorics}, 1(4):311--316, 1980.

\bibitem{liebenau2017asymptotic}
Anita Liebenau\: and Nick Wormald.
\newblock Asymptotic enumeration of graphs by degree sequence, and the degree
  sequence of a random graph.
\newblock {\em arXiv preprint arXiv:1702.08373}, 2017.

\bibitem{harary2014graphical}
Frank Harary\: and Edgar~M Palmer.
\newblock {\em Graphical enumeration}.
\newblock Elsevier, 2014.

\bibitem{BA99}
Albert-Laszlo Barabasi and Reka Albert.
\newblock Emergence of scaling in random networks.
\newblock {\em Science}, 286:509--512, 1999.

\bibitem{goyal2014sampling}
Ravi Goyal, Joseph Blitzstein, and Victor De~Gruttola.
\newblock Sampling networks from their posterior predictive distribution.
\newblock {\em Network Science}, 2(01):107--131, 2014.

\bibitem{SLH62}
S.L. Hakimi.
\newblock On realizability of a set of integers as degrees of the vertices of a
  linear graph.
\newblock {\em Journal of the Society for Industrial and Applied Mathematics},
  10:496--506, 1962.

\bibitem{VH55}
V.~Havel.
\newblock A remark on the existence of finite graphs.
\newblock {\em $\check{C}$asopis Pest. Mat.}, 80:477--480, 1955.

\bibitem{MN02}
Mark Newman.
\newblock Assortative mixing in networks.
\newblock {\em Physical Review Letters}, 89(20):208701, 2002.

\bibitem{goyal2018inference}
Ravi Goyal\: and Victor De~Gruttola.
\newblock Inference on network statistics by restricting to the network space:
  applications to sexual history data.
\newblock {\em Statistics in medicine}, 37(2):218--235, 2018.

\bibitem{qu2015effects}
Jing Qu, Sheng-Jun Wang, Marko Jusup, and Zhen Wang.
\newblock Effects of random rewiring on the degree correlation of scale-free
  networks.
\newblock {\em Scientific reports}, 5:15450, 2015.

\bibitem{MN10}
Mark~E. Newman.
\newblock {\em Networks An Introduction}.
\newblock Oxford University Press, New York, 2010.

\end{thebibliography}

\end{document}